\documentclass[aip,jcp,reprint,superscriptaddress,floatfix]{revtex4-1}

\pdfoutput=1

\usepackage{amssymb,amsmath}
\usepackage{graphicx}
\usepackage{color}

\usepackage[breaklinks]{hyperref}

\usepackage{epstopdf}

\begin{document}
\bibliographystyle{aipnum4-1}

\title{Efficiency scaling of non-coherent upconversion}
\author{Jochen Zimmermann}
\email{jochen.zimmermann@physik.uni-freiburg.de}
\affiliation{Physikalisches Institut, Albert-Ludwigs-Universit\"at Freiburg, Hermann-Herder-Str. 3, 79104 Freiburg, Germany}
\author{Roberto Mulet}
\affiliation{Physikalisches Institut, Albert-Ludwigs-Universit\"at Freiburg, Hermann-Herder-Str. 3, 79104 Freiburg, Germany}
\affiliation{Group of Complex Systems, Department of Theoretical Physics, Physics Faculty, University of Havana, Cuba}
\author{Thomas Wellens}
\affiliation{Physikalisches Institut, Albert-Ludwigs-Universit\"at Freiburg, Hermann-Herder-Str. 3, 79104 Freiburg, Germany}
\author{Gregory D. Scholes}
\affiliation{Physikalisches Institut, Albert-Ludwigs-Universit\"at Freiburg, Hermann-Herder-Str. 3, 79104 Freiburg, Germany}
\affiliation{Department of Chemistry, 80 St George Street, Institute for Optical Sciences, and Centre for Quantum Information and Quantum Control, University of Toronto, Toronto, Ontario M5S 3H6, Canada}
\author{Andreas Buchleitner}
\affiliation{Physikalisches Institut, Albert-Ludwigs-Universit\"at Freiburg, Hermann-Herder-Str. 3, 79104 Freiburg, Germany}

\begin{abstract}

A very promising approach to obtain efficient upconversion of light is the use of triplet-triplet annihilation of excitations in molecular systems.
In real materials, besides upconversion, many other physical processes take place - fluorescence, non-radiative decay, annihilation, diffusion - and compete with upconversion.
The main objective of this work is to design a \emph{proof of principle} model that can be used to shed light on the relevance of the interaction between the different physical processes that take part in these kinds of systems. Ultimately, we want to establish general principles that may guide experimentalists toward the design of materials with maximum efficiency.
Here we show, in a 1D model system, that even in the presence of these processes upconversion can be optimized by varying the ratio between the two molecular species present in this kind of materials.
We derive scaling laws for this ratio and for the maximum efficiency of upconversion, as a function of the diffusion rate J, as well as of the creation and of the decay rate of the excitations.

\end{abstract}

\maketitle

\section{Introduction}

To obtain highly efficient solar cells is one of the most important technological problems to be solved in the future. In the last years, researchers have developed new strategies to achieve this task: solar cells from novel materials, multi-junction solar cells, but also, new and cheaper fabrication techniques, etc.\cite{Brown:2009fk,Green:2009uq,Nozik:2010kx}

Recently, to overcome the efficiency limit due to spectral mismatch\cite{Shockley:1961fk}, the idea of upconverting low energy photons has been revived\cite{Ende:2009fk,Wild:2011fk,Wang:2011fk,Zou:2012ys,Cheng:2012ve,Khnayzer:2012vn}.
The idea is to absorb those photons that can not be absorbed by the solar cell, and send them back to the cell with an appropriate wavelength, ie. a wave length that can efficiently be absorbed by the cell. One example of a well studied low power upconversion mechanism is noncoherent upconversion by triplet-triplet-annihilation (TTA)\cite{Singh-Rachford:2010fk,Zhao:2011uq,Monguzzi:2012fk,Ma:2012rr,Ji:2012eu,Haefele:2012ly,Wu:2012nx,Cheng:2010fk,Merkel:2009oq,Sugunan:2009qy,Monguzzi:2008vn,Baluschev:2006uq,Laquai:2005fr}. The migration of excitation energy through assemblies or aggregates of organic light absorbing molecules is used for capturing the energy of sunlight, similar to organic solar cells and natural light harvesting systems\cite{Scholes:2011zr}, and is then transferred to a molecule where the upconversion occurs. The high absorptivity of these organic molecules, ie. in comparison to nanoparticles\cite{Zou:2012ys}, is beneficial for the upconversion process. Other techniques, such as high harmonics generation or excited state absorption, require high light intensities and usually exhibit only low upconversion quantum yield \cite{Scheps:1996vn,Zhao:2011uq}.

Successful experimental implementations of non-coherent upconversion exist for various materials \citep{Singh-Rachford:2010fk,Zhao:2011uq,Monguzzi:2012fk,Ma:2012rr,Ji:2012eu,Haefele:2012ly,Wu:2012nx,Cheng:2010fk,Merkel:2009oq,Sugunan:2009qy,Monguzzi:2008vn,Baluschev:2006uq,Laquai:2005fr}, in solutions and in thin films. To the best of our knowledge, the highest upconversion yield achieved so far is
19.3\%, reported by Ji et al. \cite{Ma:2012rr}. 
However, only few studies investigating concentration and power dependence exist\cite{Monguzzi:2012fk,Cheng:2010fk,Auckett:2009fk}, and little is known about the quantitative effect of the various loss channels on the upconversion yield.
In the present report we study a simple model system in order to address the question how the upconversion efficiency can be optimized by varying the ratio between sensitizers to acceptors -- the two molecular species that are present in materials used for upconversion by TTA (see below). We carried out model calculations to investigate the effect of sensitizer to emitter ratio, triplet lifetime and triplet creation rate on the upconversion efficiency. We find that for the optimal choice of sensitizer-to-emitter ratio the efficiency can reach almost unity, neglecting spin statistical effects. This leads us to suggest that optimal systems for non-coherent upconversion can be very efficient, when tailored to the light conditions of their environment. The results of our model give insight into dynamics of a non-coherent upconversion material and extend our intuitive understanding of the competition between different reaction pathways.

\begin{figure}[b!]
		\def\svgwidth{0.7\columnwidth}
	\begingroup%
  \makeatletter%
  \providecommand\color[2][]{%
    \errmessage{(Inkscape) Color is used for the text in Inkscape, but the package 'color.sty' is not loaded}%
    \renewcommand\color[2][]{}%
  }%
  \providecommand\transparent[1]{%
    \errmessage{(Inkscape) Transparency is used (non-zero) for the text in Inkscape, but the package 'transparent.sty' is not loaded}%
    \renewcommand\transparent[1]{}%
  }%
  \providecommand\rotatebox[2]{#2}%
  \ifx\svgwidth\undefined%
    \setlength{\unitlength}{139.625132bp}%
    \ifx\svgscale\undefined%
      \relax%
    \else%
      \setlength{\unitlength}{\unitlength * \real{\svgscale}}%
    \fi%
  \else%
    \setlength{\unitlength}{\svgwidth}%
  \fi%
  \global\let\svgwidth\undefined%
  \global\let\svgscale\undefined%
  \makeatother%
  \begin{picture}(1,0.75658493)%
    \put(0,0){\includegraphics[width=\unitlength]{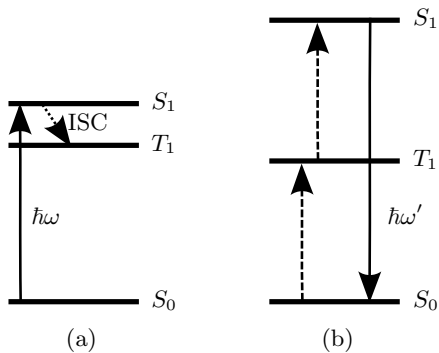}}%
    \put(0.12744126,0.00830908){\color[rgb]{0,0,0}\makebox(0,0)[lb]{\smash{(a)}}}%
    \put(0.68894476,0.00830908){\color[rgb]{0,0,0}\makebox(0,0)[lb]{\smash{(b)}}}%
    \put(0.31527838,0.09425349){\color[rgb]{0,0,0}\makebox(0,0)[lb]{\smash{$S_0$}}}%
    \put(0.31527838,0.43803114){\color[rgb]{0,0,0}\makebox(0,0)[lb]{\smash{$T_1$}}}%
    \put(0.31527838,0.52970518){\color[rgb]{0,0,0}\makebox(0,0)[lb]{\smash{$S_1$}}}%
    \put(0.05171551,0.27760157){\color[rgb]{0,0,0}\makebox(0,0)[lb]{\smash{$\hbar \omega$}}}%
    \put(0.1319303,0.48386816){\color[rgb]{0,0,0}\makebox(0,0)[lb]{\smash{ISC}}}%
    \put(0.89024747,0.09425349){\color[rgb]{0,0,0}\makebox(0,0)[lb]{\smash{$S_0$}}}%
    \put(0.89002603,0.40365343){\color[rgb]{0,0,0}\makebox(0,0)[lb]{\smash{$T_1$}}}%
    \put(0.89024747,0.71305318){\color[rgb]{0,0,0}\makebox(0,0)[lb]{\smash{$S_1$}}}%
    \put(0.83408476,0.27760164){\color[rgb]{0,0,0}\makebox(0,0)[lb]{\smash{$\hbar \omega '$}}}%
  \end{picture}%
\endgroup%
	\caption{Sketch of level structure for (a) sensitizer and (b) emitter molecules. Photonic transitions are indicated by solid lines. Dashed lines represent transitions induced by triplet energy transfer. (a) Upon photon absorption, the excitation energy $\hbar\omega$ is transferred from the singlet $S_1$ state to the triplet $T_1$ state, via intersystem crossing (ISC). This process occurs with almost unity quantum yield in molecules used for non-coherent upconversion. (b) The emitter level structure allows for excited singlet state occupation by pooling the energy of two triplets. Thus, the emitter molecule can emit a photon of energy $\hbar\omega^{\prime } > \hbar \omega $ via fluorescence.}
	\label{fig:tta_upconversion}
\end{figure}

The state of the art scheme for upconversion by TTA involves two types of molecules \cite{Singh-Rachford:2010fk,Zhao:2011uq}, see also \autoref{fig:tta_upconversion}: The \emph{sensitizer} molecules harvest solar energy. Upon photon absorption a sensitizer molecule is excited from the ground state $S_0$ to its first excited singlet state, $S_1$. This is followed by rapid intersystem crossing (ISC)\cite{KASHA:1950fk} to a triplet state, $T_1$. The $T_1$-$S_0$ transition is spin forbidden and therefore suppressed, but the transition can be mediated by spin-orbit interaction. Hence, these triplets have a relatively long lifetime, which enables a large triplet diffusion length. Eventually, the triplet is transferred to the second moiety, the \emph{emitter} molecules (sometimes also referred to as \emph{acceptors}). The emitter molecule has the property that the level spacing between $S_1$ and ground state equals approximately twice the spacing of its triplet and ground state. Therefore, it is energetically possible to pool the energy of two triplets in the emitter $S_1$ state. Thus, if an emitter is in its $T_1$ state and another triplet, from a sensitizer or emitter, is close by, the emitter's $S_1$ state can be populated. If both participating molecules are of the same species, this is referred to as homo, else as hetero process.\cite{Nickel:1993fk,Keivanidis:2009kx} A strong $S_1$-$S_0$ coupling consequently leads to emission of an upconverted photon. In fact, fluorescence from the emitter $S_1$ state is experimentally observed \cite{Baluschev:2006uq,Nickel:1993fk,Keivanidis:2009kx}.

However, other processes compete with this upconversion mechanism: A triplet-triplet encounter can not lead to upconversion if both of the two molecules are sensitizers, since the sensitizer $S_1$ state does not have the appropriate energy. Instead, if both molecules are sensitizers, the well known triplet-triplet-annihilation \cite{SUNA:1970cr,GAIDIDEI:1980bh} $T+T\rightarrow T$ occurs, which in statistical physics literature is referred to as \emph{coagulation} and has been studied extensively \cite{ben-Avraham:1998uq,Masser:2000vn,Masser:2001fk,Masser:2001kx}. Consequently, a triplet-triplet encounter can only lead to upconversion if at least one of the involved molecules is an emitter. If that is the case, and the spin states of the two triplets add to $m_S=0$, upconversion can take place: $T+T\rightarrow h \nu$.

Transport of triplets in the material can occur either by diffusion of the molecule itself, in case of solution based materials \cite{Monguzzi:2008ys}, or by triplet hopping between neighbouring molecules.
In our model system, we consider a solid state material, where transport of triplets occurs exclusively by triplet hopping which can be described by Miller-Abrahams rates\cite{Koehler:2009vn,Koehler:2011kx}, see \autoref{sec:model} below.
In total, we thus need to consider five processes to capture the full kinetics: Triplet creation in sensitizer molecules, triplet hopping between neighbouring molecules, coagulation (if two triplets meet at a sensitizer molecule), upconversion (if two triplets meet at an emitter molecule), and triplet decay via phosphorescence. In principle, it is straightforward to write down a master equation for these processes, however, solving it is a daunting task.

Similar systems were studied before, for example
on-site annihilation and fusion in one-dimensional disordered media was studied\cite{Schutz:1997vn,Schutz:1998nx} by mapping the master equation of the corresponding classical random walk onto a quantum Hamiltonian. Sch\"utz et al. \cite{Schutz:1998nx} found an approximation to the ensemble averaged time evolution of the many body random walk, by studying the many particle random walk in its dual environment, where calculations simplify for specific Hamiltonians.
In principle, it is also possible to implement coagulation, particle decay and creation in the Hamiltonian. Unfortunately, the problem no longer simplifies in the dual environment if the Hamiltonian contains particle creation.\\
Ben-Avraham et al. \cite{BENAVRAHAM:1990ys} studied particle diffusion on a one-dimensional, translationally invariant chain including coagulation and particle creation. Their work is based on the method of intervals or inter particle distribution function. This function gives the probability that a subunit of the system contains no particle, as function of the size of the subsystem. The benefit of this framework is that it gives a good intuition of the system dynamics. The latter can be described by a differential equation in the inter particle distribution function, which can be solved using standard methods \cite{BENAVRAHAM:1990ys}. In infinite systems, this method gives an exact solution. However, application of this method for other boundary conditions is not trivial. In particular, it is not possible to use it for finite systems with absorbing boundaries and particle creation: The mean density vanishes on the absorbing boundary, but is finite inside the system, due to particle creation. Hence, the mean density clearly must be a function of the spatial coordinate, and, consequently, the system is not translationally invariant and the method can not be applied. One important result - suggested already earlier \cite{KOPELMAN:1988fk}, but confirmed by Ben-Avraham et al. \cite{BENAVRAHAM:1990ys} - is that coagulation in one-dimensional systems scales with the mean density cubed. We will use this fact below.

Because of these difficulties, we will follow a more intuitive approach, looking for analytical solutions only in the regime in which the equations involved are linear. We will then deduce general scaling arguments, for the linear and nonlinear regime, from first principles. All our results will be underpinned with numerical simulations.

The remaining four sections of this paper are structured as follows: In the next section we introduce the upconversion model system. Then, we present numerical findings for the upconversion efficiency in this model. Afterwards, we perform an analytical analysis for the mean field density and the upconversion efficiency. Finally, we present an intuitive argument that results in scaling laws for a critical system parameter, and for the upconversion efficiency.

\section{The Model}
\label{sec:model}

To enhance energy transfer from sensitizer to emitter, one should choose an emitter with a triplet
energy level below that of the sensitizer triplet state.
In this case, energy transfer from emitter to sensitizer is unlikely to happen, since the
corresponding Abrahams-Miller transition rate is  exponentially suppressed for $\Delta E\gg k_B T$,
with temperature $T$ and energy difference $\Delta E\geq 0$ between the respective levels
\cite{Koehler:2009vn,Koehler:2011kx}.
This is the case for configurations where high efficiency was observed: In the experiment reported
by Cheng et al. \cite{Cheng:2010fk} rubrene was used as emitter, with a lowest triplet state energy
of $\approx 1.15eV$, and a lifetime of $\approx100\mu s$ \cite{LIU:1977qf,HERKSTROETER:1981ve}.
The sensitizer PdPQ$_4$ was excited with 670nm laser pulses, corresponding to $\approx 1.8~{\rm eV}$, ie. much higher
than the energy of the $T_1$ emitter state. Since $\Delta E=(1.8-1.15)~{\rm eV}=0.65~{\rm eV}\gg k_BT\simeq 0.03~{\rm eV}$ for room temperature, it is reasonable to neglect emitter to sensitizer energy transfer in our model.
On the other hand, we describe sensitizer to emitter transfer and transfer between two sensitizers by a constant rate $J$. This rate is determined by the intermolecular distance and the electronic coupling between the respective states, which we assume to be identical for all neighbouring molecules.
The lifetime of emitter triplets, $\approx 100\mu s$, is, when compared with other relevant
experimental values used by Cheng et al.\cite{Cheng:2010fk}, the slowest timescale in our model.
Hence, it is well justified to assume an infinite lifetime of emitter triplets in our model.
As further simplification, the diffusion limited reactions upconversion\cite{Singh-Rachford:2009rt} and coagulation are modelled as on-site interaction with infinite rate. Thus, two
triplets occupying the same site immediately undergo upconversion (coagulation), if the site is an
emitter (sensitizer), see below.
Finally, the lifetime of the sensitizer triplet depends strongly on the material, values from $0.25ns$ up to $100\mu s$ were reported\cite{Zhao:2011uq}.

As a consequence of the simplifications introduced above, excitations are trapped at emitters. Thus, in our one-dimensional model we can think of emitters as absorbing barriers. This constraint simplifies the kinetics. As we will see in \autoref{sec:analytical_results}, it allows us to reduce the study of dynamics to sensitizers only, since each triplet which escapes to an emitter will eventually contribute to emission of an upconverted photon.  
In our model, we consider only hetero type upconversion, because this type yields higher efficiencies for otherwise identical parameters, and is therefore more promising.

Our one-dimensional toy model is illustrated in the following sketch, where $E$ and $S$ represent emitter and sensitizer molecules, respectively.
\begin{equation}
\label{upconversionsystem}
\underbrace{E \: \overbrace{S \: S \ldots S \: S}^{l\,\Delta x} \: E \: S \: S \ldots S \: S \:\ldots\: E \: S \: S \ldots S \: S \:}_{L=n\,(l+1)\,\Delta x}
\end{equation}
The system of length $L$ is characterized by the intermolecular spacing $\Delta x$ and the emitter spacing $l\,\Delta x$. Thus, the parameter $l$ controls the sensitizer-to-emitter ratio. The system can be grouped into $n$ blocks, each containing only a single emitter and $l$ sensitizers. 

\begin{figure}[b!]
		\def\svgwidth{0.9\columnwidth}
	\begingroup%
  \makeatletter%
  \providecommand\color[2][]{%
    \errmessage{(Inkscape) Color is used for the text in Inkscape, but the package 'color.sty' is not loaded}%
    \renewcommand\color[2][]{}%
  }%
  \providecommand\transparent[1]{%
    \errmessage{(Inkscape) Transparency is used (non-zero) for the text in Inkscape, but the package 'transparent.sty' is not loaded}%
    \renewcommand\transparent[1]{}%
  }%
  \providecommand\rotatebox[2]{#2}%
  \ifx\svgwidth\undefined%
    \setlength{\unitlength}{204.1873655bp}%
    \ifx\svgscale\undefined%
      \relax%
    \else%
      \setlength{\unitlength}{\unitlength * \real{\svgscale}}%
    \fi%
  \else%
    \setlength{\unitlength}{\svgwidth}%
  \fi%
  \global\let\svgwidth\undefined%
  \global\let\svgscale\undefined%
  \makeatother%
  \begin{picture}(1,0.72474571)%
    \put(0,0){\includegraphics[width=\unitlength]{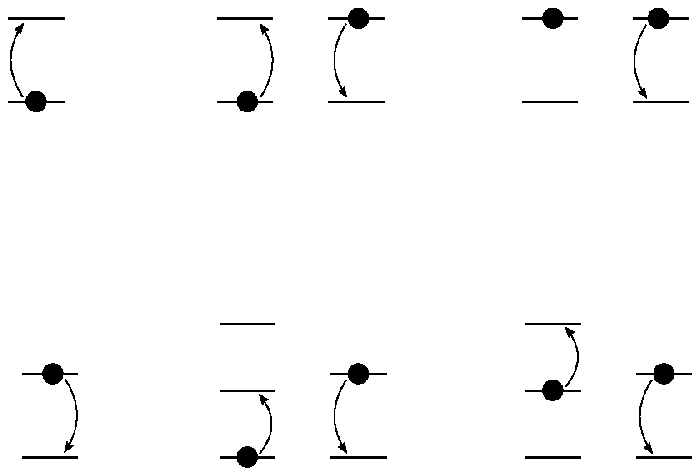}}%
    \put(0.85870121,0.00568182){\color[rgb]{0,0,0}\makebox(0,0)[lb]{\smash{(f)}}}%
    \put(0.8469473,0.51366073){\color[rgb]{0,0,0}\makebox(0,0)[lb]{\smash{(e)}}}%
    \put(0.42772449,0.00568182){\color[rgb]{0,0,0}\makebox(0,0)[lb]{\smash{(d)}}}%
    \put(0.04376342,0.51366073){\color[rgb]{0,0,0}\makebox(0,0)[lb]{\smash{(a)}}}%
    \put(0.05691662,0.00568182){\color[rgb]{0,0,0}\makebox(0,0)[lb]{\smash{(b)}}}%
    \put(0.41848926,0.51366073){\color[rgb]{0,0,0}\makebox(0,0)[lb]{\smash{(c)}}}%
    \put(0.27884163,0.08264193){\color[rgb]{0,0,0}\makebox(0,0)[lb]{\smash{$S_0$}}}%
    \put(0.27884163,0.17667322){\color[rgb]{0,0,0}\makebox(0,0)[lb]{\smash{$T_1$}}}%
    \put(0.27884163,0.2707045){\color[rgb]{0,0,0}\makebox(0,0)[lb]{\smash{$S_1$}}}%
    \put(0.57660736,0.08264193){\color[rgb]{0,0,0}\makebox(0,0)[lb]{\smash{$S_0$}}}%
    \put(0.57660736,0.20018104){\color[rgb]{0,0,0}\makebox(0,0)[lb]{\smash{$T_1$}}}%
    \put(-0.00325222,0.63899369){\color[rgb]{0,0,0}\makebox(0,0)[lb]{\smash{$\Gamma_c$}}}%
    \put(0.14255221,0.13553453){\color[rgb]{0,0,0}\makebox(0,0)[lb]{\smash{$\Gamma_d$}}}%
  \end{picture}%
\endgroup%
	\caption{Possible processes occuring in our upconversion model system. Black bars indicate the level structure, black dots represent occupied states, whereas arrows indicate transitions. Sensitizers are represented by the two levels $S_0$ and $T_1$, emitters by the three levels $S_0$, $T_1$ and $S_1$. The processes and respective rates are (a) triplet creation, with rate $\Gamma_c$, (b) triplet decay, with rate $\Gamma_d$, (c) triplet hopping, with rate $J$, (d) triplet energy transfer from sensitizer to emitter, with rate $J$, (e) triplet hopping with subsequent triplet-triplet annihilation or coagulation, with rate $J$, (f) triplet hopping with subsequent upconversion (not indicated), with rate $J$.}
	\label{fig:processes}
\end{figure}

In an ensemble of emitter and sensitizer molecules, the energy of the $T_1$ state will differ from molecule to molecule, due to dynamical disorder. This on-site energy disorder can be neglected, since for symmetric hopping rates it merely corresponds to a rescaling of the diffusion constant $D=(\Delta x)^2J$.\cite{Schutz:1998nx,ALEXANDER:1981zr,ANSHELEVICH:1982vn,ZWANZIG:1982uq} In the time average, the on-site energy differences vanish and sensitizer hopping rates can be considered symmetric. Thus, we do not expect a considerable change in our results due to this approximation. However, this rescaling will decrease the diffusion constant and, thus, the range of validity of the low density approximation decreases as well.

We perform an exact simulation of the upconversion system as depicted in \eqref{upconversionsystem}, using the aforementioned approximations and a kinetic Monte Carlo algorithm\cite{YOUNG:1966uq,BORTZ:1975vn,FICHTHORN:1991kx}. \autoref{fig:processes} lists the processes included in the simulation: (a) triplet creation by absorption of a photon by a sensitizer molecule, with rate $\Gamma_c$ - note that an excited sensitizer does not absorb photons; (b) sensitizer triplet decay, with rate $\Gamma_d$, e.g. via phosphorescence; (c) triplet hopping, with rate $J$ between two neighbouring sensitizers; (d) triplet energy transfer from sensitizer to emitter, modelled with the same rate $J$ as triplet hopping among sensitizers; (e) triplet energy transfer, with rate $J$, immediately followed by coagulation/triplet-triplet annihilation; (f) triplet energy transfer, with rate $J$, resulting in population of the emitter $S_1$ state, immediately followed by emission of an upconverted photon.

To measure the efficiency of the upconversion process, the internal quantum efficiency is a natural candidate:
\begin{equation}
\label{eq:quantumyield}
\chi_{iqe}=\frac{\text{\emph{no. of upconverted particles}}}{\text{\emph{no. of created particles}}}\text{.}
\end{equation}
However, this is a bad measure for the overall efficiency of a given configuration. It is only limited by the internal loss channels decay and coagulation. The efficiency of the absorption process is not taken into account. For example, consider a system which consists of only a single absorbing site, and of traps (emitters) otherwise. This system has $\chi_{iqe}=1$, but will absorb only a tiny fraction of the available incident intensity. 

On the other hand, the external quantum efficiency, EQE, also takes the efficiency of absorption into account,
\begin{equation}
\label{eq:upconversionefficiency_tw}
EQE=\frac{\text{\emph{no. of upconverted particles}}}{\text{\emph{no. of incident photons}}}=n_{abs}\chi_{eqe}\text{,}
\end{equation}
where the constant $n_{abs}=\sigma_{abs}(\Delta x)^2$, with sensitizer absorption cross section $\sigma_{abs}$, gives the number of created particles per photon incident in an area $(\Delta x)^2$ around a sensitizer molecule, and the efficiency measure $\chi_{eqe}$ is defined below. 
Since the sensitizers cover only a fraction $l/(l+1)$ of the total sample, we obtain:
\begin{align}
EQE\propto\chi_{eqe}
 &=\frac{\text{\emph{no. of upconverted particles}}}{\Gamma_c t\:L/\Delta x}\\
\label{eq:upconversionefficiency}
 &=\chi_{iqe}\frac{l}{l+1}\text{,}
\end{align}
where the second equality strictly speaking only holds for low densities. This assumption is fulfilled within the parameter space of consideration, ie. $J\gg\Gamma_c,\Gamma_d$, as can be seen in \autoref{fig:density}, where the density is on the order of $10^{-4} \ll 1$.
The external quantum efficiency $\chi=\chi_{eqe}$ is used from here on as efficiency measure.

\section{Numerical Results}
In our numerical studies, we simulate a chain of sensitizer and emitter molecules on a lattice structure. The dynamic processes in the simulation are restricted to those shown in \autoref{fig:processes}. Thus, dynamics in the sublattice of sensitizers is simulated exactly, whereas dynamics of emitter molecules is subject to the constraints introduced above: an infinite triplet lifetime and no energy transfer from emitter to sensitizer.
Due to these approximations, the kinetics only depends on the number of neighbouring sensitizers enclosed by emitters, but not on the number of neighbouring emitters. Therefore, we do not allow emitters to be placed next to each other on the lattice. To ease the analysis of numerical results, we study a chain of sensitizers where we place emitters equally spaced.
The nonlinear two particle interaction coagulation is implemented by limiting the number of triplets to one per lattice site. Each triplet that reaches an emitter molecule, where it will stay until another triplet arrives and upconversion occurs, contributes to the upconverted fraction.

\begin{figure}[h]
\includegraphics{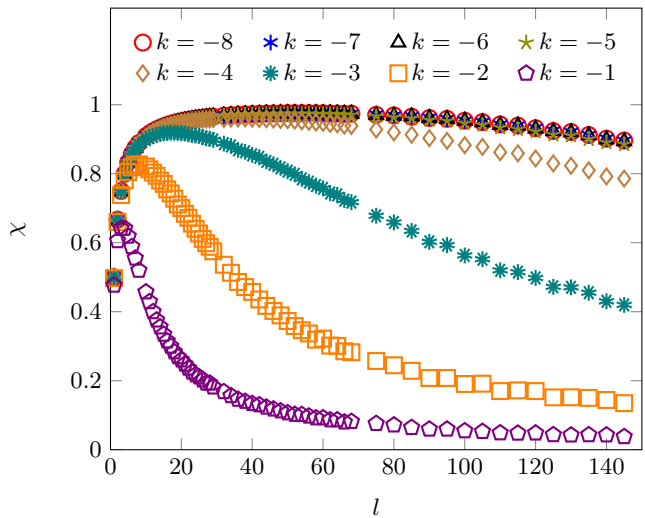}
\caption[labelInTOC]{ Numerically obtained values for the external quantum efficiency, as function of emitter spacing l, for the creation rate $\Gamma_c=10^{-6}$ (in units of the hopping rate, which is set to $J=1s^{-1}$ throughout the paper, see \eqref{eq:scaling_of_timescales}). Each curve corresponds to a decay rate in the interval $\Gamma_d=10^{k}$, $k=-1,\ldots,-8$. For decay rates with $k=-5,\ldots,-8$, the curves almost coincide. Each curve exhibits a critical value $l_c$, where the external quantum efficiency is maximized.}
\label{fig:chi_vs_la}
\end{figure}
\begin{figure}[h]
\includegraphics{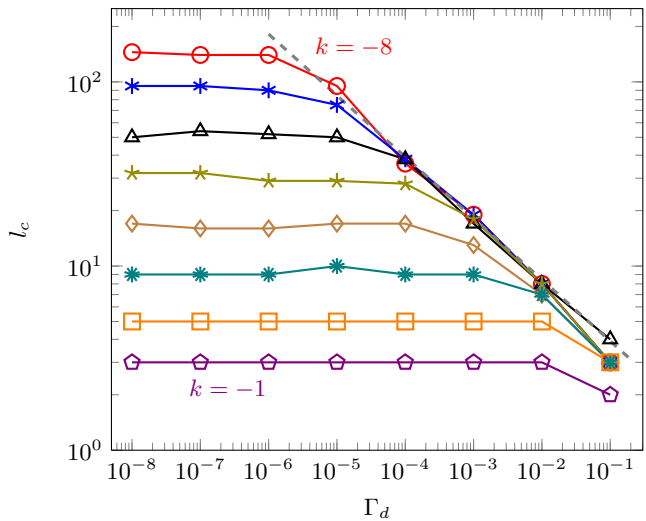}
\caption[labelInTOC]{ Numerical results for the critical value $l_c$ of the emitter spacing (ie. the value for which $\chi$ is maximal), as function of decay rate $\Gamma_d$, for creation rates $\Gamma_c=10^{k}$, $k=-1,\ldots,-8$. The connecting lines are merely a visual aid.
The two limiting regimes are clearly distinguishable: In the coagulation limited case, $l_c$ is constant, whereas the slope of $l_c$ is similar for all values of $\Gamma_c$, in the decay limited case, and coincides with the analytical prediction (dashed line) given by \eqref{eq:lc_ana}.
}
\label{fig:lac_vs_Gd}
\end{figure}

In all simulations, the rate of triplet hopping to the left and to the right was set to $J=1s^{-1}$.
The actual energy transfer rates, as given by the corresponding Miller-Abrahams rate, can be mapped to $J=1s^{-1}$ by scaling all rates in our model ($\Gamma_c$, $\Gamma_d$) with the same factor, since the final results only depend on the ratio $J/\Gamma_c$ and $J/\Gamma_d$. The scaling is described by the following equation, where $J_{true}$, $\Gamma_{d,true}$, $\Gamma_{c,true}$ refer to actual experimental values and $\tau$ is the scaling constant:
\begin{equation}
\label{eq:scaling_of_timescales}
\tau=\frac{J_{true}}{J}=\frac{\Gamma_{d,true}}{\Gamma_d}=\frac{\Gamma_{c,true}}{\Gamma_c}
\end{equation}

According to \autoref{fig:chi_vs_la}, the efficiency can vary strongly with the emitter spacing $l$, which controls the sensitizer-to-emitter ratio, see \eqref{upconversionsystem}, and is easy to control in experiments. It is therefore an ideal candidate for optimization.
For large decay rates ($\Gamma_d =10^{k}s^{-1}$, with $k=-1,-2,-3$ in \autoref{fig:chi_vs_la}), the efficiency has a pronounced peak at a very well defined value of $l$, which we name $l_c$. Decreasing the decay rates increases $l_c$. Finally, for very low decay rates, ie. $k>4$ in \autoref{fig:chi_vs_la}, the value of $l$ at which the efficiency is maximal does no longer change. $l_c$ saturates due to particle loss by coagulation, and the efficiency is near-optimal in a large interval of $l$.
This can be observed in \autoref{fig:lac_vs_Gd}, where we plot $l_c$ as a function of the decay rate $\Gamma_d$.
It becomes apparent that the behavior of $l_c$ can be classified into two different regimes: on the one hand, for large $\Gamma_c$ and small $\Gamma_d$, $l_c$ is constant as a function of $\Gamma_d$, and mainly depends on the creation rate $\Gamma_c$. In this regime, the efficiency is limited by coagulation, as shown below. On the other hand, for small $\Gamma_c$ and large $\Gamma_d$, $l_c$ is independent of $\Gamma_c$ and scales according to a power law with $\Gamma_d$, consistent with \eqref{eq:lc_ana} derived below. In this regime, the efficiency is limited by decay.
The transition between the two regimes occurs at the point in parameter space where the timescales of coagulation and decay are of the same order. We discuss this in section 5.

\begin{figure}[t]
  \includegraphics{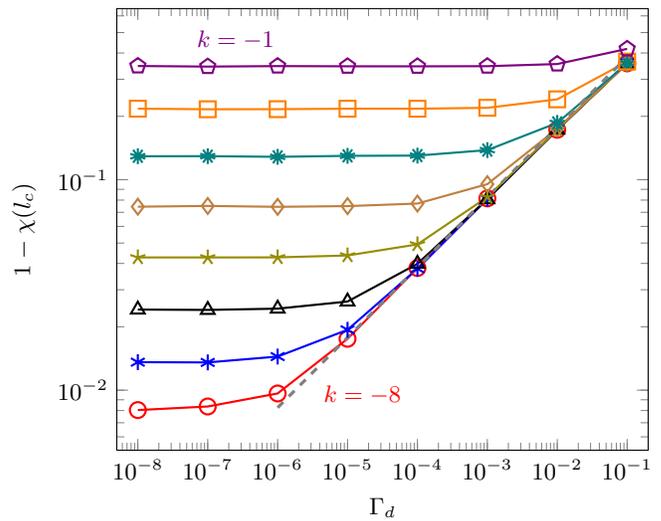}
 \caption[labelInTOC]{ Maximal value of external quantum efficiency $\chi(l_c)$ for optimal emitter spacing, as function of decay rate $\Gamma_d$, for $\Gamma_c=10^{k}$, $k=-1,\ldots,-8$. The connecting lines are merely a visual aid. For small $\Gamma_d$ and constant $\Gamma_c$, the efficiency is purely coagulation limited. For large $\Gamma_d$ it approaches the analytical prediction (dashed line) given by \eqref{eq:chic_ana}.}
 \label{fig:chic_vs_Gd}
  \end{figure}

In \autoref{fig:chic_vs_Gd} we plot the maximal upconversion efficiency $\chi(l_c)$. In order to highlight the regime of high efficiencies , we plot $1-\chi(l_c)$ on a logarithmic scale. Again, we  clearly distinguish between two different regimes: either the efficiency is constant as a function of $\Gamma_d$ and thus purely determined by $\Gamma_c$ (coagulation limited regime), or the efficiency is mainly determined by $\Gamma_d$ and approximately given by \eqref{eq:chic_ana} derived below (decay limited regime).

\section{Analytical Results}
\label{sec:analytical_results}

In the continuous limit $l\gg 1$ (i.e. spacing $\Delta x$ between sensitizers much smaller than spacing $l\Delta x$ between emitters), our model can approximately be described by the following mean field equation \cite{GAIDIDEI:1980bh,BENAVRAHAM:1990ys}
\begin{equation}
\label{eq:rateequation}
\frac{\partial \phi}{\partial t}=\Gamma_c(1-\phi)-\Gamma_d \phi+J \frac{\partial^2 \phi}{\partial x^2}-\alpha_{coag} \phi^3\text{,}
\end{equation}
where $\phi$ is the density of triplets in the sensitizer, such that $\phi dx$ is the probability to find a triplet in $dx$.
The first term on the right hand side describes creation of triplets. $\Gamma_c$ describes the rate of triplet creation at a single molecule, where the molecule spacing is $\Delta x$. In the continuous system the creation rate is thus $\Gamma_c/\Delta x$. For simplicity, we set $\Delta x=1$ (which is possible since the final result for the efficiency does not depend on the choice of $\Delta x$).
The factor $1-\phi$ accounts for the fact that only a sensitizer in its ground state can absorb a photon, which subsequently results in a triplet, whereas a sensitizer that is already in its $T_1$ state can not absorb another photon. In principle, if the sensitizer has appropriate energy levels, one has to allow for photon absorption from the excited $T_1$ state. However, the absorbed energy is immediately lost, due to coagulation, and is therefore irrelevant for the kinetics.
The second term describes triplet decay, e.g. via phosphorescence or non-radiative decay channels. The third term represents triplet diffusion with diffusion constant $D=J(\Delta x)^2\equiv J$ for $\Delta x=1$. Finally, the last term accounts for coagulation.
\eqref{eq:rateequation} is almost identical to the upconversion rate equation that is commonly used to describe triplet-triplet annihilation\cite{GAIDIDEI:1980bh}. The only difference comes from the term cubic in the density, which correctly describes annihilation in one dimension \cite{BENAVRAHAM:1990ys,KOPELMAN:1988fk}. The absorbing boundary conditions translate to
\begin{equation}
\label{eq:boundarycondition}
\phi(x_e)=0\text{,}
\end{equation}
where $x_e$ is the spatial coordinate of any emitter.

\begin{figure}[tb!]
  \includegraphics{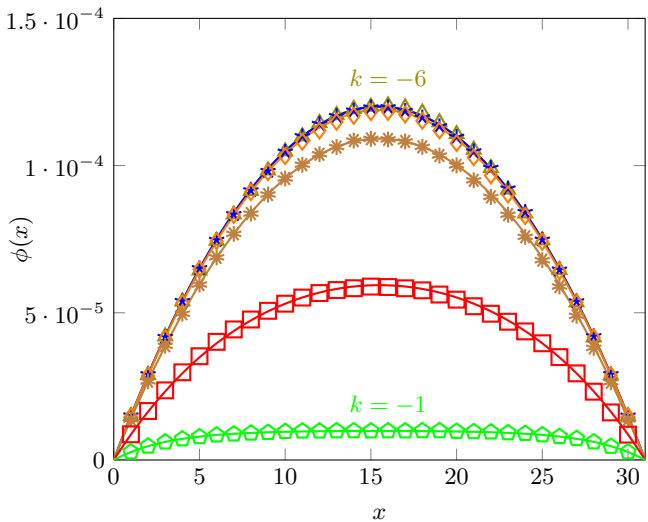}
  \caption[labelinTOC]{
  Spatial dependence of mean triplet density for creation rate $\Gamma_c=10^{-6}s^{-1}$ and emitter spacing $l=30$.
  Solid lines represent analytical results, whereas numerical data are indicated by symbols. The decay rate $\Gamma_d=10^{k}s^{-1}$ is changed for each curve, with $k=-1,\ldots,-6$. The undermost curve corresponds to $k=-1$, and the topmost to $k=-6$.  }
  \label{fig:density}
\end{figure}

A closed analytical solution to the nonlinear differential equation \eqref{eq:rateequation} is not available. However, in the low density limit, $\phi\ll 1$ or $l\Gamma_c \ll \Gamma_d+J/l^2$, the nonlinear term can be neglected, resulting in the linear differential equation
\begin{equation}
\label{eq:lineardiffequation}
\frac{\partial \phi}{\partial t}=\Gamma_c(1-\phi)-\Gamma_d \phi+J \frac{\partial^2 \phi}{\partial x^2}\text{.}
\end{equation}
For convenience, we consider only positions between the first two emitters, i.e. $0<x<(l+1)\Delta x$, since the solution of \eqref{eq:lineardiffequation} will be identical between all other emitters.
In the steady state, \eqref{eq:lineardiffequation} can be solved exactly for boundary condition \eqref{eq:boundarycondition}, resulting in
\begin{equation}
\label{eq:meandensity}
\phi(x)=\frac{\Gamma_c}{\Gamma_c+\Gamma_d}\left[1-\frac{\cosh\left[\alpha(l+1-2x/\Delta x)\right]}{\cosh\left[\alpha(l+1)\right]}\right]\text{,}
\end{equation}
with $\alpha=\sqrt{(\Gamma_c+\Gamma_d)/J}$.
In \autoref{fig:density}, this result is compared with numerical data obtained from the exact many-particle simulation as described in \autoref{sec:model}. Not surprisingly, the prediction of the analytical equation is correct in the low density regime $\phi\ll 1$, which is realized for all curves shown in \autoref{fig:density}. Remarkably, \eqref{eq:meandensity} also shows a good agreement with numerics for $\Gamma_c=10^{-6}s^{-1}$, $\Gamma_d=10^{-6}s^{-1}$, which, according to \autoref{fig:lac_vs_Gd}, is in the coagulation limited regime. This shows that, also at small densities, the efficiency may be limited by coagulation -- provided that the decay rate is even smaller. In this case, the efficiency is close to 1, see \autoref{fig:chic_vs_Gd}.

The flux through the boundaries and the number of created particles define the internal quantum efficiency. According to Fick's law, the flux through the boundary is given by $J$ times the first derivative of $\phi$, hence
\begin{equation}
\label{eq:discreteinternalquantumefficiency}
\begin{split}
\chi_{iqe}&=\left.2J \frac{\partial\phi}{\partial x}\right|_{1/2} \left(\int_{1/2}^{(l+1/2)}dx\:\Gamma_c(1-\phi)\right)^{-1}\\
&=\frac{4\alpha\sinh[\alpha l]}{\cosh[\alpha(l+1)]\Gamma_d l +\sinh[\alpha l]\Gamma_c/\alpha}\text{,}
\end{split}
\end{equation}
where the factor $2$ takes the two boundaries into account, with $\alpha$ as above. The integrand corresponds to particle creation, see \eqref{eq:rateequation}, and the limits $1/2$ and $l+1/2$ of integration, with corresponding evaluation of the flux at $x=1/2$, ensure that we count exactly $l$ sensitizer molecules, and not $l+1$.
Due to the neglect of coagulation, the above treatment should correctly reproduce only the behaviour in the decay limited regime. To check this, we evaluate the maxima of the external quantum efficiency $\chi=l/(l+1)\chi_{iqe}$, with $\chi_{iqe}$ given by \eqref{eq:discreteinternalquantumefficiency}, under the assumption  $\Gamma_d\gg\Gamma_c$. Under the additional assumption $\Gamma_d\ll J$ (i.e., that triplet hopping is much faster than decay), we obtain the following analytical expression 
\begin{equation}
\label{eq:lc_ana}
l_c^{(d)}= \left(\frac{6J}{\Gamma_d}\right)^{1/3}\text{,}
\end{equation}
for the optimal emitter spacing in the decay limited regime.
Using \eqref{eq:discreteinternalquantumefficiency} and \eqref{eq:upconversionefficiency}, it is easy to verify that:
\begin{equation}
\label{eq:chic_ana}
1-\chi(l_c)^{(d)} = \left(\frac{9\Gamma_d}{16J}\right)^{1/3} \text{.}
\end{equation}
As expected, both expressions agree with the numerical data shown in \autoref{fig:lac_vs_Gd} and \autoref{fig:chic_vs_Gd} (dashed lines).

\section{Scaling Analysis}

Above, we already derived the scaling behaviour in the low density limit. Since there is no analytical steady state solution of \eqref{eq:rateequation} for the high density limit available, we present a more intuitive argument in this section, which works in both, low and high density regimes.

Besides triplet escape to the emitter and subsequent upconversion, there are two other triplet loss mechanisms in the system: Decay, which governs the low density regime, and coagulation, which is predominant in the nonlinear high density regime, see \autoref{fig:lac_vs_Gd}. According to the diffusion equation, a triplet placed in the center of the system needs the time $\tau_e=l^2/J$ till it reaches the boundary. This defines the timescale of escape. The decay timescale is $\tau_d=1/\Gamma_d$. Coagulation occurs on the timescale in which a triplet is created in the system with length $l$, thus $\tau_{co}=1/(l\Gamma_c)$. If a particle exists already in the system, there will be two particles after a time $\tau_{co}$ on average, and, consequently, coagulation can occur. The transition between decay and annihilation limited regimes is expected to take place when the respective timescales are of the same order:
\begin{equation}
\label{eq:regimetransition}
\frac{\tau_d}{\tau_{co}}=\frac{l\Gamma_c}{\Gamma_d}=1\text{.}
\end{equation} 

Clearly, the efficiency must depend on how the three relevant timescales mentioned above compare to each other.
On the other hand, we have seen that the only relevant parameters that define the optimal structure are the creation rate in the coagulation limited regime, and the decay rate in the decay limited regime. Thus, in each regime, the fraction of triplets which escape and are upconverted is determined by only two timescales, which are $\tau_e$ and either $\tau_d$ or $\tau_{co}$. Therefore, in the steady state the effective particle creation rate equals the sum of the two relevant loss rates:
\begin{equation}
\Gamma_c(1-\phi)=\Gamma_e+\Gamma_{d,co}\text{.}
\end{equation}
The internal quantum efficiency is identical to the quotient of the escape rate and the effective rate with which particles are created. This implies: 
\begin{equation}
\chi_{iqe}=\frac{\Gamma_e}{\Gamma_e+\Gamma_{d,co}}=\frac{\tau_{d,co}}{\tau_e+\tau_{d,co}}\text{.}
\end{equation}
In order to obtain an efficient system we demand $\tau_e < \tau_{d,co}$. Expanding in the small parameter up to first order, we find
\begin{equation}
\label{eq:scalingargument}
\chi=\frac{l}{l+1}\left(1-\frac{\tau_e}{\tau_{d,co}}\right)\text{.}
\end{equation}
As we will show below, this intuitive equation yields the correct scaling behaviour for optimal emitter spacing and optimal external quantum efficiency, when analysed in the respective regimes. By evaluating the maxima of \eqref{eq:scalingargument}, we obtain the optimal emitter spacing $l_c$. Solution of
\begin{equation}
\frac{\partial \chi}{\partial l}=0
\end{equation}
for $l$, evaluated in the limit of $\Gamma_c+\Gamma_d\ll J$, yields $l_c$ and its scaling behaviour.
\begin{figure}[tb!]
\includegraphics{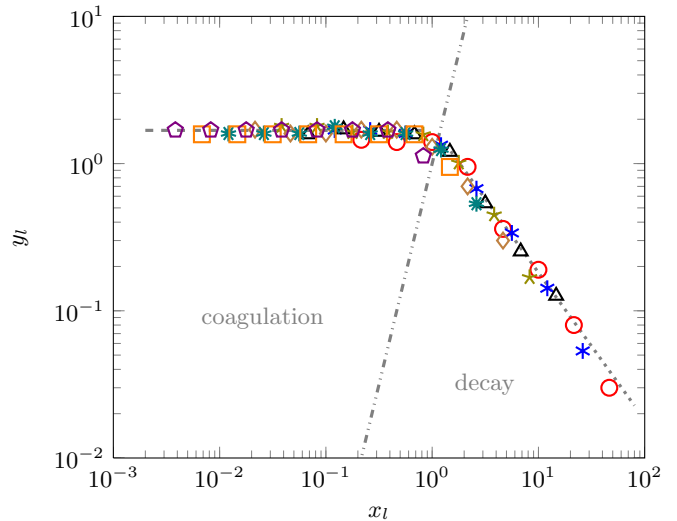}
\caption[labelInTOC]{ Optimal value of scaled emitter spacing $y_l=l_c\Gamma_c^{1/4}$ as function of the scaled variable $x_l$, see equations \eqref{eq:x_l} and \eqref{eq:y_l}, for different values of $\Gamma_c$ (labels as in \autoref{fig:lac_vs_Gd}). Due to the rescaling, the data for different $\Gamma_c$ collapse onto a single curve. Therefore, the scaled optimal emitter spacing is completely characterized by the single variable $x_l$. The dashed/dotted lines mark
the behaviour $y_l^{(co)}=6^{1/3}/x_l$ and $y_l^{(d)}=8^{1/4}$
in the coagulation and decay limited regime, respectively. The dash-dotted line separates decay and coagulation limited regimes in parameter space, see \eqref{eq:regimetransition} and \eqref{eq:coag_decay_sep}.}
\label{fig:lac_scaled}
\end{figure}

\begin{figure}[b!t]
 \includegraphics{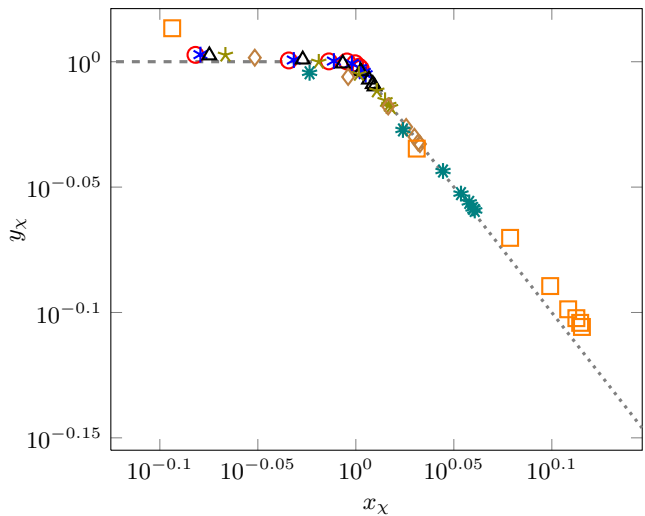}
  \caption[labelInTOC]{Scaled maximal efficiency $y_{\chi}$, see \eqref{eq:ychi}, as a function of the scaled variable $x_{\chi}$, see \eqref{eq:xchi}. Labels encode different values of $\Gamma_c$, as in \autoref{fig:lac_vs_Gd}. The dashed/dotted lines mark the scaling prediction \eqref{eq:y_chi_coagulation}/\eqref{eq:y_chi_decay} in the coagulation/decay limited regime, respectively. The numerical data for different $\Gamma_c$ and $\Gamma_d$ collapse onto a single curve, except for very large $\Gamma_c$ or $\Gamma_d$, where the assumption $\Gamma_c+\Gamma_d\ll J$ fails, ie. $x_{\chi}\lesssim10^{-0.1}$ and $x_{\chi}\gtrsim10^{0.07}$. The transition between decay and coagulation limited regime of the optimal efficiency occurs at $x_{\chi}=1$, where $x_{\chi}<1$ corresponds to decay limited optimal efficiencies.}
\label{fig:chic_scaled}
\end{figure}
In the decay limited low density regime we find
\begin{equation}
\label{eq:scaling_lac_d}
l^{(d)}_c \propto \left(\frac{J}{\Gamma_d}\right)^{1/3} 
\end{equation}
for the scaling of $l_c$, and by substitution in \eqref{eq:scalingargument} we obtain the scaling law for the efficiency:
\begin{equation}
\label{eq:scaling_chic_d}
1-\chi(l^{(d)}_c) \propto \left(\frac{\Gamma_d}{J}\right)^{1/3}\text{.}
\end{equation}
The scaling laws \eqref{eq:scaling_lac_d} and \eqref{eq:scaling_chic_d} agree with the analytical results \eqref{eq:lc_ana} and \eqref{eq:chic_ana}. However, it is important to note that, while \eqref{eq:scalingargument} gives the correct scaling behaviour, it can not be used to quantitatively predict absolute values. The reason is the following: For evaluation of \eqref{eq:scalingargument} we insert an escape rate, which is only an upper bound. In fact, the escape rate depends on the spatial coordinate, since a particle sitting close to the boundary will escape faster than a particle sitting in the center of sensitizer molecules. Averaging over all spatial coordinates would, in principle, yield the correct value. However, for averaging one has to know the density in order to give each contribution the correct weight, and the density in the coagulation limited regime is not available analytically. Thus, the absolute value of the inserted escape rate is not correct, but the scaling of $l_c$ with the system parameters is, as it yields the correct exponent and agrees with the numerical data.

Similarly, we obtain scaling laws in the coagulation limited high density regime, where an exact analytical result is not available. We find
\begin{equation}
l^{(co)}_c \propto \left(\frac{J}{\Gamma_c}\right)^{1/4}
\label{eq:scaling_lac_co}
\end{equation}
for the scaling of the optimal emitter spacing, whereas the maximal efficiency scales as 
\begin{equation}
1-\chi(l^{(co)}_c) \propto \left(\frac{\Gamma_c}{J}\right)^{1/4}\text{.}
\label{eq:scaling_chic_co}
\end{equation}
As above, we do not inherit the correct prefactors from \eqref{eq:scalingargument}. 
In order to determine these factors, we define rescaled variables
\begin{equation}
\label{eq:x_l}
x_l=\frac{\Gamma_d^{1/3}}{\Gamma_c^{1/4}}
\end{equation}
for the rates $\Gamma_c$ and $\Gamma_d$
\begin{equation}
\label{eq:y_l}
y_l=l_c\,\Gamma_c^{1/4}\text{.}
\end{equation}
for the optimal emitter spacing.
According to equations \eqref{eq:lc_ana} and \eqref{eq:scaling_lac_co}, the rescaled emitter spacing
$y_l$ as a function of $x_l$ is constant in the coagulation limited regime, and given by $y_l^{(d)}=(6J)^{1/3}/x_l$ in the decay limited regime.
Moreover, according to \eqref{eq:regimetransition}, the border separating the decay limited regime from the coagulation limited regime
turns into
\begin{equation}
\label{eq:coag_decay_sep}
y_l=x_l^3\text{,}
\end{equation}  
such that $x_l<y_l^{1/3}$ corresponds to the coagulation limited regime, and
$x_l>y_l^{1/3}$  to the decay limited regime. These predictions are confirmed by \autoref{fig:lac_scaled}, where we show the same data
as in \autoref{fig:lac_vs_Gd} (optimal emitter spacing as a function of $\Gamma_d$) -- but now expressed in terms of the rescaled
variables $x_l$ and $y_l$. Indeed, all data points in \autoref{fig:lac_scaled}  collapse onto a single universal curve,
which, on the double logarithmic scale, is given by two straight lines corresponding to the two regimes discussed above.
Furthermore, one of the lines  agrees with the above prediction $y_l^{(d)}=6^{1/3}/x_l$ for the decay limited regime,
whereas the value of the constant $y_l^{(c)}=8^{1/4}\simeq 1.7$ observed for the coagulation limited regime allows to determine the missing prefactor in \eqref{eq:scaling_lac_co} as
\begin{equation}
l^{(co)}_c = \left(\frac{
8J}{\Gamma_c}\right)^{1/4}\text{.}
\label{eq:lac_scaling_annihil}
\end{equation}
Concerning the coagulation limited efficiency, see \eqref{eq:scaling_chic_co}, a fit of the numerical data
shown in \autoref{fig:chic_vs_Gd} (for $\Gamma_d\to 0$) yields the prefactor:
\begin{equation}
\label{eq:chic_scaling_annihil}
1-\chi(l_c^{(co)})=\left(\frac{10\Gamma_c}{33
J}\right)^{1/4}\text{.}
\end{equation}
Correspondingly, if we define scaled variables for $\chi$ as:
\begin{equation}
x_{\chi}=\left(1-\left(\frac{9\Gamma_d}{16J}\right)^{1/3}\right)\left(1-\left(\frac{10\Gamma_c}{33J}\right)^{1/4}\right)^{-1}\text{,}
\label{eq:xchi}
\end{equation}
and
\begin{equation}
y_{\chi}=\chi(l_c)\left(1-\left(\frac{9\Gamma_d}{16J}\right)^{1/3}\right)^{-1}\text{,}
\label{eq:ychi}
\end{equation}
the rescaled maximal efficiency as given by \eqref{eq:chic_ana} and \eqref{eq:chic_scaling_annihil} is found as
\begin{equation}
\label{eq:y_chi_decay}
y_{\chi}^{(d)}=
1\text{,}
\end{equation}
in the decay limited regime, whereas
\begin{equation}
\label{eq:y_chi_coagulation}
y_{\chi}^{(co)}=
\frac{1}{x_{\chi}}\text{,}
\end{equation}
in the coagulation limited regime. This is confirmed by  \autoref{fig:chic_scaled}, where we plot the same data as in
\autoref{fig:chic_vs_Gd} in terms of the scaled variables $x_\chi$ and $y_\chi$.
A collapse onto a single universal curve can also be observed here. Note that some numerical values slightly deviate from the predicted curve, since these data points are not strictly in the limit $\Gamma_d+\Gamma_c\ll J$, for which the scaling equations were derived.

\section{Conclusions}
Optimization of the sensitizer-to-emitter ratio can enhance the upconversion efficiency drastically. We identified two characteristic kinetic regimes of the system, one where single particle decay limits the efficiency, and a second one where the nonlinear particle interaction coagulation (triplet-triplet annihilation) is the limiting factor. For each regime, we derived analytic expressions for the optimal ratio of sensitizers and emitters. The provided scaling laws for the optimal structure and efficiency reveal the universal behaviour of our model system. In future work, we would like to answer the question how the scaling changes in higher dimensions. Knowledge of the scaling laws of a 3D system, which most likely differ from the 1D scaling laws shown here, will make it simpler to verify the suitability of a given material and optimize it for non-coherent upconversion.\\

\begin{acknowledgments}
This work was supported by a grant from the Ministry of Science and the Arts of Baden-Wuerttemberg (Az: 33-7532.20/694),
together with the Wissenschaftliche Gesellschaft in Freiburg im Breisgau, to JZ, GDS, and AB. RM acknowledges financial support from the Alexander von Humboldt foundation.
\end{acknowledgments}

\end{document}